\documentstyle[a4,epsf,psfig]{article}
\begin{document}
\title{Density waves and $1/f$ density fluctuations in  granular flow}
\author{Gongwen Peng \thanks{Permanent address: Institute of Physics,
    Academia  Sinica, Beijing, China.} and Hans J. Herrmann \\ 
H\"ochstleistungsrechenzentrum, Forschungszentrum J\"ulich, \\ D--52425
J\"ulich, Germany
%HLRZ, KFA J\"ulich, D--52425 J\"ulich, Germany
}
\date{\today} 
\maketitle 
\begin{abstract}
  \noindent We simulate the granular flow in a narrow pipe with a
  lattice--gas automaton model. We find that the density in the system
  is characterized by two features. One is that spontaneous density
  waves propagate through the system with well--defined shapes and
  velocities. The other is that density waves are so distributed to
  make the power spectra of density fluctuations as $1/f^{\alpha}$
  noise. Three important parameters make these features observable and
  they are energy dissipation, average density and the rougness of the
  pipe walls.  \\ 
\end{abstract}

\noindent
%\raisebox{17.5cm}[][]{preprint HLRZ 11/94}
\newpage
\begin{center}
\bf I. INTRODUCTION
\end{center}

Granular materials, like powders or beads, are widely processed in
industry. In this kind of materials, many unusual phenomena such as
size segregation \cite{Williams, Haff, Rosato, Devillard}, heap
formation and convection cells under vibration \cite{Farady, Evesque,
  Taguchi, Gallas}, and anomalous sound propagation \cite{Liu, Jaeger}
have been found. Even in simple geometries like hoppers and pipes,
their flow under gravity still shows complex dynamics \cite{nature,
  Baxter, Poschel}. Experiments \cite{nature, Baxter, Poschel} and
molecular dynamics (MD) simulations \cite{Poschel, Ristow, Jysoo} show
that the granular particles do not flow uniformly but rather form
density waves (or shock waves). The density fluctuation in the
granular flow was observed to follow a $1/f^{\alpha}$ noise both in
experiments \cite{nature, Baxter, peter} and in MD simulations
\cite{Ristow}.\\ 

We have studied the granular flow through a narrow pipe using a
lattice--gas automaton (LGA).  Since a general theory for granular
media is not yet available, people have used various methods to get
better understanding about the complicated rheological behavior of
granular media. Among the different methods are MD \cite{Haff, Ristow,
  shearflow, LeeHans}, Monte Carlo simulations \cite{Devillard, MC},
event driven algorithms \cite{luding} and cellular automaton
\cite{CA}.  So far the most widely used method is MD \cite{MDbook}
which simulates the granular materials on a ``microscopic'' level (the
grain's level). MD has been recognized to be very successful in
simulating granular materials.  MD needs, however, much computer time
to give reasonable results. The same situation was also faced in
classical fluid mechanics some years ago when Frisch, Hasslacher and
Pomeau \cite{FHP} proposed lattice--gas automata as an alternative to
the direct solution of the equation of motion. The basic idea behind
LGA is that a properly defined cellular automaton with appropriate
conservation laws should lead to the Navier--Stokes equation which is
nothing but an expression of the conservation of momentum.  Guided by
this spirit we have designed an LGA for granular flow. Some of our
results have been briefly reported in Ref. \cite{us}.  In the present
paper we will give a detailed account of our results obtained with
this model. The paper is organized as follows. We describe the model
in the following section and present the numerical results in section
III. A discussion is contained in section IV. \\ 
\begin{center}
  \bf{II. THE LGA MODEL}
\end{center}

We consider an LGA at integer time steps $t=0,1,2,\cdots$ with $N$
particles located at the sites of a two--dimensional triangular
lattice which is $L$ sites long vertically and $W$ sites wide
horizontally. Gravity is parallel to one of the lattice axes.
Periodic boundary conditions are used in the vertical direction while
fixed boundary conditions are set for the walls. At each site there
are seven Boolean states which refer to the velocities, $\vec{v}_i
(i=0,1,2,\cdots 6)$. Here $\vec{v}_i (i=1,2,\cdots 6)$ are the nearest
neighboring (NN) lattice vectors and $\vec{v}_0=\vec{0}$ refers to the
rest (zero velocity) state. Each state can be either empty or occupied
by a single particle. Therefore, the number of particles per site has
a maximal value of $7$ and a minimal value of 0. The time evolution of
the LGA consists of a collision step and a propagation step. In the
collision step particles change their velocities due to collisions and
in the subsequent propagation step particles move in the directions of
their velocities to the NN sites where they collide again.\\ 

The system is updated in parallel. Only the specified collisions shown
in figure 1 can deviate the trajectories of particles. All collisions
conserve mass and momentum.\\ 

For two-- and three--body collisions, we have the probabilistic rules
shown in figure 1 (a). The probablity that a configuration may take
place is shown next to the configuration. If the parameter $p$ is
nonzero, it means that energy can be dissipated in the collision. \\ 

The collision rules for moving particles with a rest particle involve
typical mechanisms of granular flow.  Intuitively one can understand
them as follows. Rest particles in a region will decrease the local
granular temperature which is defined as the (kinetic) energy, causing
a decrease in pressure in that region. The resulting pressure gradient
will lead to a migration of particles into that region, increasing its
density and decreasing its pressure and granular temperature even more
\cite{Savage, Goldhirsch}. That means that rest particles will
effectively attract rest particles nearby. However, due to the
restriction of LGA that the rest state at one site can at most be
occupied by one particle, we must introduce some additional
constraints in our model. For example, two moving particles colliding
with a rest particle from opposite directions can stop each other in
accordance with momentum conservation. But on each site only one rest
particle is allowed.  Therefore, the two particles stay at rest on the
NN sites where they originally came from.  However, on these sites
there may already exist other rest particles.  To make things simple,
we will still use the on--site collision as defined and temporarily
allow more than one rest particle on a site during the collision.
Immediately after the collision, the extra rest particles randomly hop
to NN sites until they find a suitable site with no rest particle
already sitting there.  Only in this way can we incorporate the
mechanisms mentioned above.  The collision rules with rest particles
are shown in figure 1 (b).\\ 

The driving force of the flow is gravity. We simply incorporate its
effect as follows. A rest particle decides to have a velocity along
the direction of gravity with probability $g$, if the resulting state
is empty at that time. Any moving particle can change its velocity by
a unit vector along gravity with probability $g$, if the resulting
state is possible on the trianglar lattice used. These are depicted in
figure 1 (c). \\ 

The sites at the walls of the system only have two directions into
which the particles can move. So, a particle colliding with the wall
from one direction can be bounced back with probability $b$ and
specularly reflected into the other direction with probability $1-b$.
If $b=0$, the walls are smooth (perfect no--slip condition).
Otherwise, the walls have some roughness. The collision rule is
depicted in figure 1~(d).\\ 
\begin{center}
  \bf{III. RESULTS}
\\
A. Density Contrast
\end{center}

%\indent
We evolve the system according to the collision rules defined above.
The initial configuration of the system is set to be random in the
sense that every state (except the rest state) of each site is
randomly occupied according to a preassigned average density $\rho$.
Figure 2 shows the time evolution of the density in the pipe during
the early stage for $p=0.1, g=0.5, b=0.5$. Here the system has length
$L = 2200$ and width $W = 11$ with average density $\rho=1.0$ (note
the range of $\rho$ is between $0$ and $7$).  We divided the pipe
along the vertical direction into $220$ bins with equal length of $10$
(total length $L=2200$) and counted the number of particles $n_i$ in
the $i$th bin, i. e., within an area $W \times 10$. Therefore the
spatial distribution of the density along the pipe is represented by a
one--dimensional array $\{n_i, i=1,2,\cdots\}$. Figure 2 (a) shows the
plots of $n_i$ for 9 successive snapshots every 2000 time steps. Time
increases upward and the direction of gravity is from left to right.
We see that at $t=0$ the distribution of density of the initial
configuration is just a white noise having no structure (the lowest
curve in figure 2 (a)). As time developes, a wave is formed,
travelling in the pipe along the direction of gravity. In figure 2 (b)
we use a gray scale to represent the time evolution of the density
distribution.  We plot $n_i$ in the $i$th bin by a grayscale which is
a linear function of $n_i$.  The $ n_i (i=1,2,\cdots)$ at a given time
are plotted from left to right while the densities at different time
steps are plotted from bottom to top as time increases. Gravity is
from left to right. We see that initially the density is rather
uniform and gradually regions of high density are being formed out of
the homogeneous system. A high density region may also die out and two
high density regions may merge to form a single one.  It seems
possible that these are the same density waves (or shock waves) which
were also observed in experiments \cite{Baxter, Poschel} and MD
simulations \cite{Poschel, Ristow, Jysoo}.\\ 

The density contrast can be characterized by the following quantity:
\begin{equation}
C(t) = \frac{1}{\bar{n}} \sum_{i=1}^{K} {|n_i(t)-n_{i+1}(t)|}
\end{equation}
where K is the number of bins we divided the pipe into and $\bar{n}$
is the average number of particles in a bin. The periodic boundary
condition ensures $n_{K+1}=n_1$. In figure 3 (a) we plot the density
contrast $C(t)$ versus time step $t$. This curve was obtained by
averaging over 64 simulations for systems with length $L=256$, width
$W=11$.  It is clear that the density contrast increases from zero
before it saturates at a fixed value, indicating that the density
waves are being formed spontaneously from the uniform initial
configurations.  In figure 3 (b) the relexation of the kinetic energy
$E(t)$ of the system is plotted. We take the kinetic energy of one
particle to be unity. Since the particles have a kinetic energy of
either one (for moving particles) or zero (for rest particles) in our
LGA model, the loss of kinetic energy due to dissipation is equal to
the increase in the number of rest particles. We see from figure 3 (b)
that the system loses its kinetic energy in the early stage and then
reaches a steady state where the kinetic energy loss due to
dissipation is compensated by the work of gravity (i.e. the potential
energy loss). \\ 

%\newpage
\begin{center}
  \bf{B. Density Profile}\\ 
\end{center}

%\indent
From figure 2 we know that there is a strongest density wave which is
quite different from the rest.  To obtain the shape of this density
wave, we did many simulations and averaged. For each system size, we
run $64$ simulations. For each simulation run, we recorded the density
field at each time step in the steady state to keep $2048$ density
fields. The density fields are then shifted vertically so that they
overlap each other maximally. Since the density wave travels along the
pipe, this shifted distance should be equal to the time interval of
the two density fields multipied by the average velocity in that time
interval. We use this shift distance to determine the velocity of the
density wave in the next subsection.  Here the maximal overlap rule is
applied hierarchically to obtain a clearer shape of the density wave.
$64$ simulation runs are then averaged to give the final density
profiles which are presented in figure 4. We notice that the density
wave has a non--symmetric shape and its wave front is sharper than the
backside of the wave.  The width of the wave also has a scaling
relation with the system length (almost linear) and the amplitude of
the wave only increases a little bit as the system length increases. A
similar density profile was produced in Ref. \cite{flekkoy} with a
lattice Boltzmann model.
\\ 
\begin{center}
\bf{C. Density Wave Velocity}\\
\end{center}

%\indent
As mentioned above, the velocity of the density wave can be measured
by the distance shifted along the pipe to make maximal overlap. If one
divides the shifted distance by the time interval, one will obtain the
average velocity in this time interval. However, we can not be $a
\hspace{3pt} priori$ sure that the velocity is constant all the time
steps.  So, alternatively we first chose a reference scheme and then
distance and time are measured from this reference scheme. The
velocity was measured by plotting the shifted distance versus the
shifted time.  In figure 5 such a distance--versus--time plot is
shown. This plot is obtained by averaging 64 simulation runs each
consisting of 2048 density fields. The linearity of this curve
indicates that the density wave propagate along the pipe with a
constant velocity. The velocity is the slope of the line.  This is a
real--space determination of the velocity. In the next subsection we
will give another method of measuring the velocity which is a
by--product of the Fourier transformation. In the following we
determine all the velocities by this Fourier transformation method and
we have checked that the results given by the two methods coincide.\\ 
\begin{center}
\bf{D. $1/f$ noise in the density fluctuation}\\
\end{center}
%\indent

To characterize the density fluctuations in a certain region with
time, we calculate their power spectra.  We recorded the number of
particles in a bin. The LGA is performed for very long time steps so
that we obtain good statistics to analyse each power spectrum. We
first substract the mean value from the data, otherwise there would be
a huge peak at $f=0$ in the power spectra. We calculated the spectra
using a standard FFT routine. The power is essentially the square of
the amplitude of the Fourier Transformation of a time series. But to
get better statistics, average process has been used. We broke the
time series into $S$ segments of $M$ points each. On each segment an
FFT was performed using a Parzen window \cite{book} and the powers of
the resulting spectra were averaged.  We used $S=4$ and $M=16384$ for
most of the results.  One representative power spectrum is shown in
figure 6 for systems with $ g=0.5, b=0.5, \rho=1.0$ and $p=0.8$.  In
this figure we observe a sharp peak. This peak is due to the
contribution of the strongest density wave observed in figure 2. The
frequency of this peak corresponds to a wave velocity of $Lf/T_0$
where $L$ is the pipe length and $T_0$ is the time interval of
recording (we recorded the data every $T_0 = 10$ time steps). The
velocity measured in this way coincides with the direct measurement in
real space (see above subsection).  Apart from this peak one sees that
there is a background having a power law behaviour where the spectrum
falls off as $1/f^\alpha$. The exponent $\alpha$ is found to be around
1.33 for the parameters used in figure 6. The power--law decay in the
power spectra was also observed in experiments \cite{Baxter, nature,
  peter}. In the following subsections we will show how the exponent
and the velocity depend on the parameters $(p, \rho, g$, and $b)$. \\ 
\begin{center}
\bf{E. White Noise}\\
\end{center} 
%\indent

As we reported in \cite{us}, both dissipation and the roughness of
walls are the neccessary conditions for the presence of travelling
density waves. To see whether the $1/f$ noise is associated to the
density waves, we also performed the power spectra for systems without
dissipation ($p=0$) and sytems without roughness on the walls ($b=0$).
These results are shown in figure 7 where we see white noise. The
large peak in the spectrum for system without roughness on the walls
is due to the fact that particles are perfectly reflected on the
walls, generating a wave--like motion of density along the pipe (the
velocity of this peak is exactly $1/2$ which is a geometric effect of
the lattice used). White noise is also experimentally observed when
the walls are smooth \cite{peter}. Together with our earlier results
\cite{us}, figure 7 shows that the $1/f^{\alpha}$ noise is associated
with the density waves. \\ 
\begin{center}
\bf{F. Dependence on Dissipation}\\
\end{center} 
%\indent

From above we know that for systems without dissipation the power
spectra are just white noise, i.e, $\alpha=0$. How does the exponent
$\alpha$ change with the dissiption parameter $p$ of our model? In
figure 8 (a) we show the dependence of $\alpha$ on $p$. Each exponent
in this figure was extracted from the average power spectra of 32
simulation runs, thus ensuring acceptable error bars. The other
parameters which are kept fixed are respectively: $\rho=1.0$, $b=0.5$,
$g=0.5$. From figure 8 we see that the exponent of the power--law
decay in the spectra has approximately a constant value provided that
there is dissipation in the granular system. Without dissipation,
$\alpha$ would be zero. The exponent jumps to a constant value when
$p$ changes from zero to a non-zero value.  From this point of view,
this figure reinforces the idea that the mere existence of dissipation
can give rise to a significant change in the physics of the system
even if the degree of dissipation is minute \cite{Goldhirsch, us}. In
Ref. \cite{us} we provided another explaination to this idea, i.e.,
the density waves disappear when the system has no dissipation.\\ 

Figure 8 (b) shows that the velocity of the density wave does not
change with dissipation within the error bars. The parameters used for
the model and the averaging are the same as for figure 8 (a). \\ 

Using a Kolmogorov--Obukhov approach revised for space--intermittent
systems, \mbox{Bershadskii} \cite{Isarei} proposed that the exponent
which we found to be around $1.33$ for our model \cite{us} might be an
universal value of $4/3$ for scalar fluctuations convected by
stochastic velocity fields in dissipative systems. Our numerical
results for the present LGA model show that the exponent does not
change with $p$, $b$ and $g$ (see the following subsections), but the
average density $\rho$ does affect the exponent and this will be
discussed in the next subsection.\\ 
\begin{center}
\bf{G. Dependence on  Average Density}\\
\end{center}
%\indent

In figure 9 (a) we show how the exponent depends on the average
density $\rho$. For very low density ($\rho \leq 0.8$) the exponent is
zero, thus the density fluctuation in the pipe is just white noise.
Since in this density region the interaction among the particles is so
weak that no collective motion can be formed, the exponent can be
easily understood. For average densities above $1.0$, the exponent
increases with $\rho$. We found $\alpha \approx 1.86$ when $\rho=2.0$.
We did not go beyond $\rho=2.0$ since our LGA model becomes less valid
as the average density becomes higher.  In the present model we
introduced a rest state which can be occupied only by one particle.
Therefore, the model is not valid when the number of rest particles
exceedes the total number of sites. When the average density increases
high enough, the number of rest particles due to the fixed dissipation
rate $p$ might be too large so that the model becomes less valid for
granular flow.\\ 

The velocity of the density wave is constant when the average density
changes. This is shown in figure 9 (b).\\ 
\begin{center}
\bf{H. Dependence on Boundary Roughness and  Gravity}\\
\end{center}

As we noted in Ref. \cite{us} and discussed in subsection E, the
roughness of the walls is essential to the formation of density waves.
Without roughness, no density waves propagates in the system. We
calculated the power spectra for such cases and found that the density
fluctuation is white noise and the exponent $\alpha=0$. When the
roughness parameter $b$ is turned on even if b is very small, the
power spectrum changes. The exponents $\alpha$ are constant for any
non--zero value of $b$.  This phenomenon is illustrated in figure 10
(a). Changing gravity magnitude we found no change in the exponent.\\ 

In figure 10 (b) we present two curves to show how the velocity of the
density wave varies with the roughness of the walls and the magnitude
of gravity. The upper curve shows that the velocity decreases a little
bit as $b$ increases from $0.25$ to $1.0$. This small difference is
due to the fact that the applied gravity is large enough to dominate
the velocity. As the magnitude of gravity becomes smaller the velocity
is more sensitive to the roughness of walls as shown in the lower
curve of figure 10 (b).  These two curves also show that the velocity
changes with gravity.  All these results are reasonable to our daily
experienece. For experimentist, to change the magnitude of gravity can
be performed by putting the system on an oblique desk instead of
letting the pipe vertical.\\ 
\begin{center}
\bf{I. Open Systems}
\end{center}
%\indent

From the experimental point of view, open boundary condition is more
realistic than the periodic boundary condition. Here we consider open
systems.  The LGA for open systems is defined in analogy with the
model described in section II, with the exception that the periodic
boundary condition in the vertical direction is replaced by an open
boundary condition.  Initially the pipe is empty. Particles are then
injected from the upper boundary by a constant rate $I$ and leave the
system at the lower boundary without coming back. The injection rate
is defined as follows.  On each site of the upper boundary we consider
the states whose corresponding velocities point into the system. If
such a state is not occupied, it can be filled with probability of
$I$.  A time--evolution of the density in the pipe is shown in figure
11 (a). Densities at a given time are plotted from left to right while
densities at different time are plotted from bottom to top as time
increases. Gravity acts from left to right. From this plot we observe
that high density regions can also be formed in open systems. These
high density regions may travel along the pipe until they leave the
system from the lower border or they may die out during their
propagation. There are also more than one high density regions at one
time, in contrast to what we observed in figure 2 in periodic systems.
It seems to us that all the density waves travel with a constant
velocity in figure 11 (a). So we measure this velocity using
density--density correlation function which is defined as:
\begin{equation}
C(R,T) = \sum_{i,t}^{}{n_i(t) n_{i+R}(t+T)}
\end{equation}
where $n_i(t)$ is the number of particles in the {\em i}'th bin at
time step $t$. In figure 11 (b) we plot the correlation function
against the time difference $T$ for a fixed spatial separation $R=30$.
An observable peak exists at $ T_c = 880$ in the correlation function
which gives the velocity of the density waves $v = R\delta/T_c = 0.34$
where $\delta = 10$ is the length of a bin.
\\ 

We have measured the density fluctuation in a bin very close to the
bottom of the pipe. Its power spectrum is found to follow $1/f$ noise
only around a critical injection rate $I_c = 0.52$ and white noise
otherwise.  Figure 12 (a) shows three power spectra for different
injection rates, $I = I_c$, $I < I_c$ and $I > I_c$. The power
spectrum for $I=I_c$ falls off with a slope close to -1 in the
log--log plot. The exponent $\alpha$ for the power spectra
$1/f^{\alpha}$ is shown in figure 12 (b) for different injection rates
$I$.  It seems that the power spectrum is $1/f$ noise only at the
critical point. The critical injection rate is found to be independent
of the model parameters and the system sizes. We guess it might be
dependent on the lattice (here trianglar lattice is used). \\ 

To identify what the critical injection rate means, we investigate the
nature of the two phases seperated by the critical injection rate. We
find that in the phase for $I > I_c$ the system is clogging while for
$I < I_c$ particles pass the system freely without clogging. Figure 12
(c) shows the total number of particles in the system for different
injection rates. It is clear that the total number of particles
increases with time in the very early stage for every injection rate.
This is due to the fact that we simulate the process starting from an
empty pipe. After this relaxation time, the system reaches a steady
state for $I < I_c$ where the number of incoming particles is on
average equal to that of outgoing particles, thus keeping the total
number of particles in the system constant. However, for $I = 0.54$
which is slightly larger than $I_c = 0.52$, the total number of
particles increases all the time, meaning that some particles
accumulate in the system. Thus, the phase for $I > I_c$ can be
identified as a clogging phase.  It is therefore seems that $I_c$ is
the maximal injection rate that the system can sustain without
clogging. In Ref. \cite{wolfgang} Verm\"ohlen $et \hspace{2pt} al$
also found a critical inflow rate into a hopper with the
time--to--clog diverging at the critical point.
\\

\begin{center}
\bf IV. Disscussion
\end{center}

Using a simple lattice--gas automaton model we have shown that density
waves can be formed either from uniform initial conditions in the
periodic case or by injecting particles into an open system. Energy
dissipation is an important factor for this kind of instability.
Goldhirsch and Zanetti \cite{Goldhirsch} have shown that a gas
composed of dissipative particles is unstable to the formation of high
density clusters. They have observed clusters of high density in a
system without an external field (like gravity). Here we observe high
density regions travelling in the system under the action of gravity.
Density waves were also observed in traffic jam models \cite{Jams}
which might bear some relevance to granular systems.\\ 

The density fluctuations in our systems are found to follow
$1/f^{\alpha}$ noise with $ 0 \leq \alpha \leq 2$. Power--law spectra
have also been observed in experiments \cite{nature, Baxter, peter}
and MD simulations \cite{Ristow}. It is clear from our numerical
results that $1/f^{\alpha}$ spectra with $\alpha \neq 0$ are
associated with the propagation of density waves.  In the simulation
of open systems, we find a critical injection rate. At the critical
point the system has its maximal throughput without clogging. In
experiments one usually has a hopper above the pipe to ensure constant
refilling. Particles flow into the pipe by the action of gravity and
in fact particles flow down at the maximum rate. That is to say, the
system has its maximal outflow, which means the system self--organizes
into the critical state at $I_c$.  At the critical state the density
fluctuation follows a $1/f$ noise. The explanation of the ubiquitous
$1/f$ noise in granular flow is open.  In the traffic jam model, Nagel
\cite{Kai} also found the state of maximal throughput to be critical.
\\

We note here that the construction of the LGA model is not unique.
K\'arolyi and Kert\'esz \cite{Janos} have independently designed an
LGA model where the rest particles are located in the bonds in
addition to the rest particles on the sites. The use of LGA models to
study other phenomena in granular materials (e.g., the pile of sand or
convection under vibration) is in progress \cite{Herrmann}. \\ 
\vspace{1cm}

We thank Stephan Melin, Cristian Moukarzel, Thorsten P\"oschel, Stefan
Schwarzer, Hans-J\"urgen Tillemans and Wolfgang Verm\"ohlen for useful
discussions.\\

\newpage
\noindent
{\bf Figure Captions}\\

Figure 1: (a) Probabilistic collision rules for two-- and three--body
collisions. Thin arrows represent particles and small circles stand
for rest particles. The number next to a configuration is the
probability that the configuration takes place; (b) Collision rules
for moving particles with a rest particle. Immediately after the
collision, more than one rest particle on a site will hop to the
nearest neighbouring sites randomly until they find a suitable site
with no rest particle already there. (c) Gravity may change the
momentum of the particle by a unit vector in the direction of gravity.
(d) Collision rule for a moving particle colliding with the wall.\\ 

Figure 2: Time evolution of the density $n_i \{i=1,2,\cdots 220\}$
divide in $220$ bins along the pipe of $L=2200$, $W=11$ and
$\rho=1.0$.  Densities at a given time are plotted from left to right
(direction of gravity) while densities at different time steps are
plotted from bottom to top (direction of time increase). (a) 9
successive snapshots every 2,000 time steps from $t=0$; (b)
Time--evolution every 80 time steps during $40,000$ time steps.  The
grayscale of each bin is a linear function of $n_i$. Darker regions
correspond to higher densities. \\ 

Figure 3: (a) The density contrast C(t) versus time step t. (b)
Kinetic energy loss $E(0)-E(t)$ versus time step t. Here E(0) is the
kinetic energy at t=0. The kinetic energy of a moving particle is
chosen as energy unit.\\ 

Figure 4: Density as a function of position X along the pipe.  The
average has been made in the perpendicular direction. The model
parameters are $\rho=1.0$, $p=0.1$, $g=0.5$, $b=0.5$. The width is
fixed for various pipe lengths, $W=11$. \\ 

Figure 5: Real--space determination of the velocity of density wave.
The horizontal axis is the time interval while the vertical axis is
the displacement of the wave obtained by maximal overlap. The velocity
is the slope of the line which is $0.36 \pm 0.05$ for a system with
$L=512$, $W=11$, $\rho=1.0$, $p=0.1$, $g=0.5$, $b=0.5$.\\ 

Figure 6: Power spectrum $P(f)$ of the time series of the density
fluctuation inside a region in a pipe of length L=220 and width W=11.
The model parameters are $p=0.8$, $b=0.5$, $g=0.5$, $\rho=1.0$.  The
time series of the density fluctuation were recorded every 10 time
steps and the time period correponding to a frequency $f$ is $10/f$.\\ 

Figure 7: Power spectra $P(f)$ of the time series of the density
fluctuation inside a region in a pipe of length L=220 and width W=11.
Either without dissipation or with smooth walls, white noise is
observed. The model parameters for the system without dissipation are
$p=0$, $b=0.5$, $g=0.5$, $\rho=1.0$ while $p=0.5$, $b=0$, $g=0.5$,
$\rho=1.0$ for the system with smooth walls.\\ 

Figure 8: Dependence on the dissipation parameter $p$. The model
parameters kept fixed are $b=0.5$, $g=0.5$ and $\rho=1.0$. (a) The
power--law decay exponent $\alpha$ of the power spectra. (b) The
velocity of the density wave.\\ 

Figure 9: Dependence on the average density $\rho$. The model
parameters kept fixed are $p=0.5$, $b=0.5$ and $g=0.5$. (a) The
power--law decay exponent $\alpha$ of the power spectra. (b) The
velocity of the density wave.\\ 

Figure 10: Dependence on the roughness $b$ of the walls. (a)
Power--law decay exponent $\alpha$ in the power spectra. The model
parameters kept fixed are $p=0.5$, $g=0.5$ and $\rho=1.0$. (b)
Velocity of the density wave for two different magnitudes of gravity.
Here $p=0.5$ and $\rho=1.0$. \\ 

Figure 11: (a) Time evolution of the density $n_i \{i=1,2,\cdots
220\}$ in the $100$ bins in the pipe of $L=1000$, $W=5$ and $I = 0.5$.
Other model parameters are $p=0.5$, $b=0.5$, $g=0.2$.  Densities at a
given time are plotted from left to right (direction of gravity) while
densities at different time steps are plotted from bottom to top
(direction of time increase).  Time goes from $0$ to $40,000$ time
steps. The grayscale of each bin is a linear function of $n_i$. Darker
regions correspond to higher densities. (b) Two--point
density--density correlation function $C(R,T)$ versus time difference
$T$ at a fixed spatial separation $R=30$ for the evolution shown in
(a). \\

Figure 12: (a) Three typical power spectra for different injection
rates $I$, $I=I_c=0.52, I < I_c, I > I_c$. The model parameters kept
fixed are $p=0.5$, $b=0.5$, $g=0.5$. The two lower curves have been
shifted vertically for clarity.  (b) The exponent $\alpha$ in the
power spectra $1/f^{\alpha}$ for different injection rates. (c) Total
number of particles in the system $N(t)$ versus time step $t$ for
different injection rates.  Each curve is an average over $32$
simulations. \\ 

\newpage
\begin{figure}[p]
\centerline{
\epsfxsize=6.0cm
\epsfbox{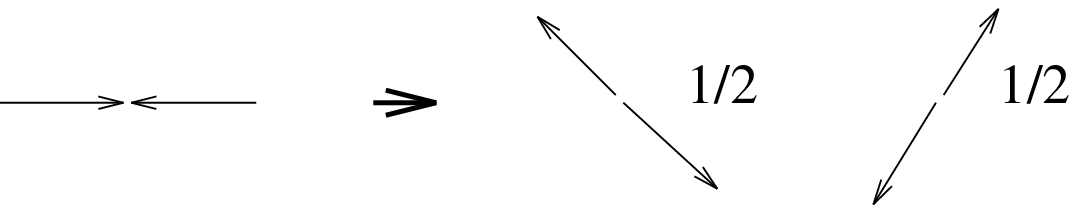}}
\vspace{1.0cm}
\centerline{
\epsfxsize=6.0cm
\epsfbox{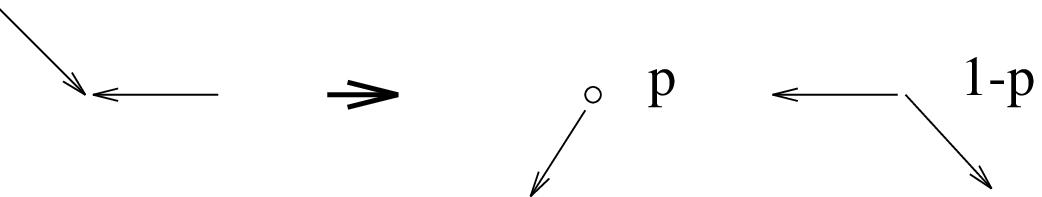}}
\vspace{1.0cm}
\centerline{
\epsfxsize=6.0cm
\epsfbox{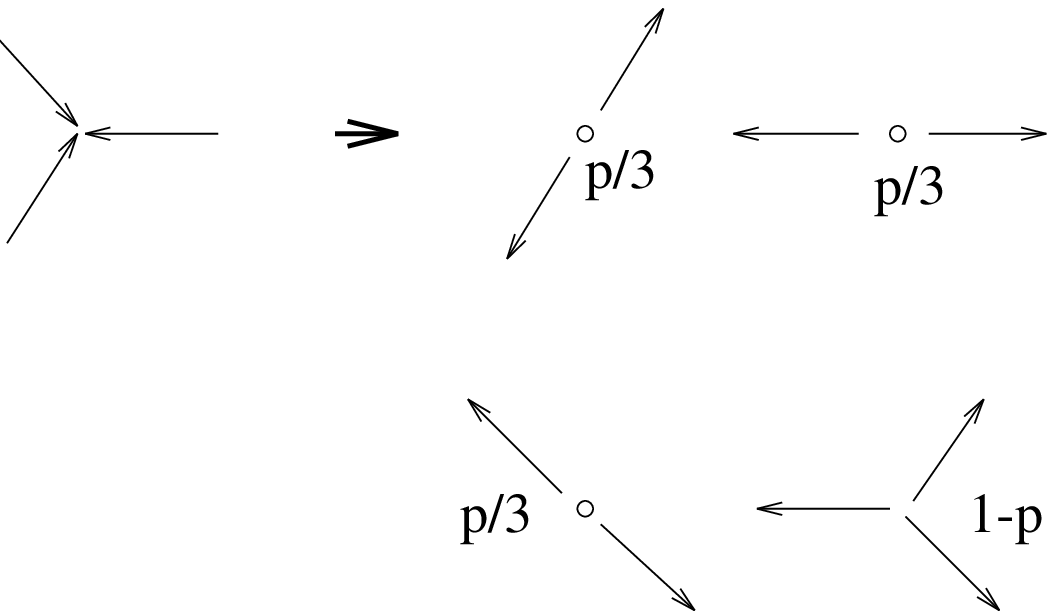}}
\centerline{(a)}
\vspace{1.5cm}
\centerline{
\epsfxsize=4cm
\epsfbox{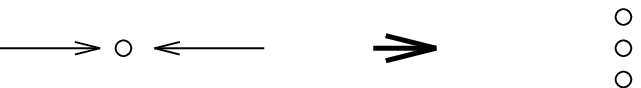}}
\vspace{1.0cm}
\centerline{
\epsfxsize=4cm
\epsfbox{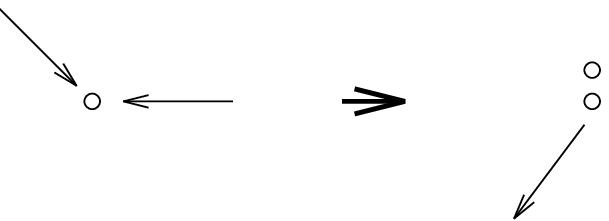}}
\centerline{(b)}
\end{figure}
\begin{figure}[p]
\centerline{
\epsfxsize=2cm
\epsfbox{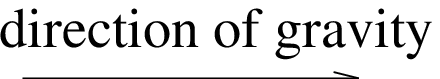}}
\vspace{0.8cm}
\centerline{
\epsfxsize=6.0cm
\epsfbox{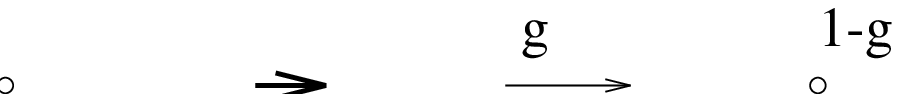}}
\vspace{1.0cm}
\centerline{
\epsfxsize=6.0cm
\epsfbox{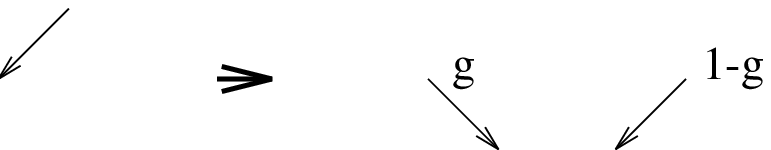}}
\vspace{1.0cm}
\centerline{
\epsfxsize=6.0cm
\epsfbox{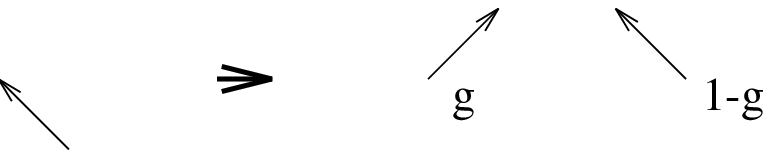}}
\centerline{(c)}
\vspace{1.5cm}
\centerline{
\epsfxsize=6.0cm
\epsfbox{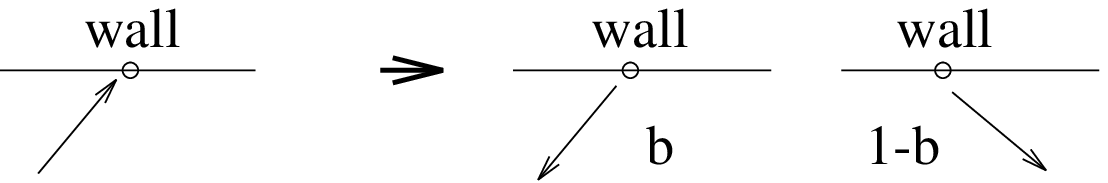}}
\centerline{(d)}
\caption{ (a) Probabilistic collision rules for two-- and
  three--body collisions. Thin arrows represent  particles
  and  small circles stand for rest particles. The number next to a
  configuration is the probability that the configuration   takes
  place; (b) Collision rules for moving particles with a
  rest particle. Immediately after the collision, more than one rest
  particle on a site will hop to the nearest neighbouring sites
  randomly until they find a suitable site with no rest particle
  already there. (c) Gravity may change the momentum of the
  particle by a unit vector in the direction of gravity. (d) Collision
  rule for a moving particle colliding
  with the wall.} 
\end{figure}

\begin{figure}[p]
\centerline{
\epsfxsize=4.0cm
\epsfbox{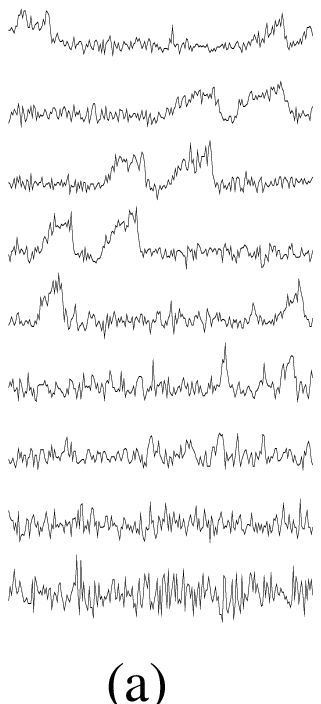}
\hfil
\epsfxsize=6.0cm
\epsfbox{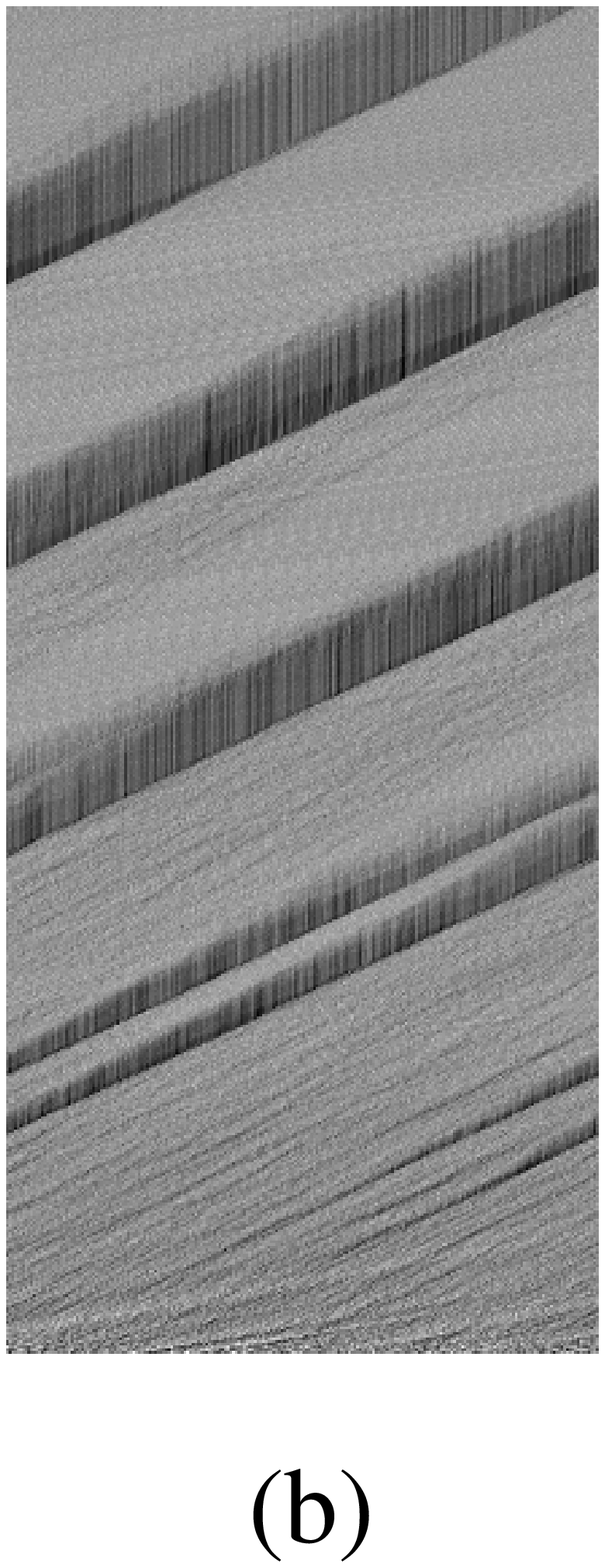}}
\caption{ Time evolution of the density $n_i \{i=1,2,\cdots 220\}$ 
divide in  $220$ bins along the pipe of $L=2200$, $W=11$ and $\rho=1.0$.
 Densities at a given time are
plotted from left to right (direction of gravity) while
densities  at different time steps are plotted from bottom to top
 (direction of time increase). (a) 9 successive snapshots every 2,000
 time steps from $t=0$; (b) Time--evolution  
every 80 time steps during 
$40,000$ time steps. 
The  grayscale of each bin is a linear function of $n_i$. Darker
regions  correspond to higher densities.  }
\end{figure}

\begin{figure}[e]
\vspace{5.9cm}
\includegraphics{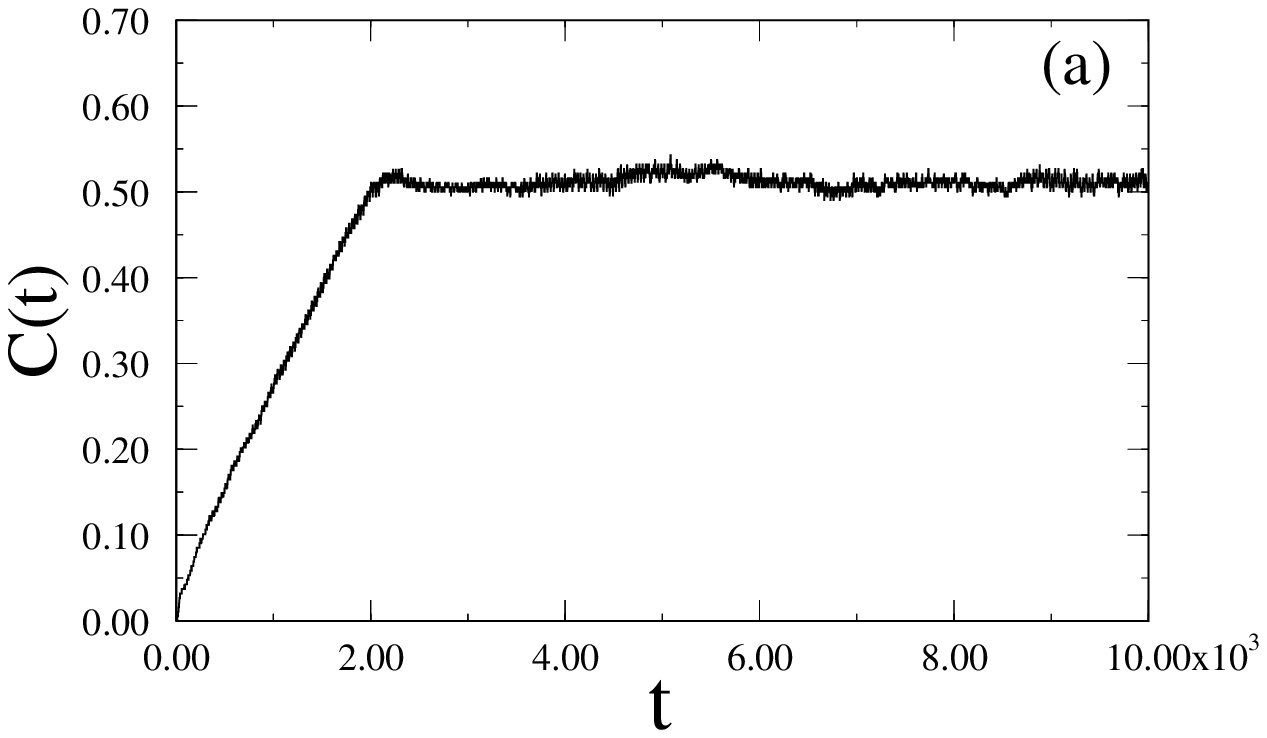}
\vspace{6.9cm}
\includegraphics{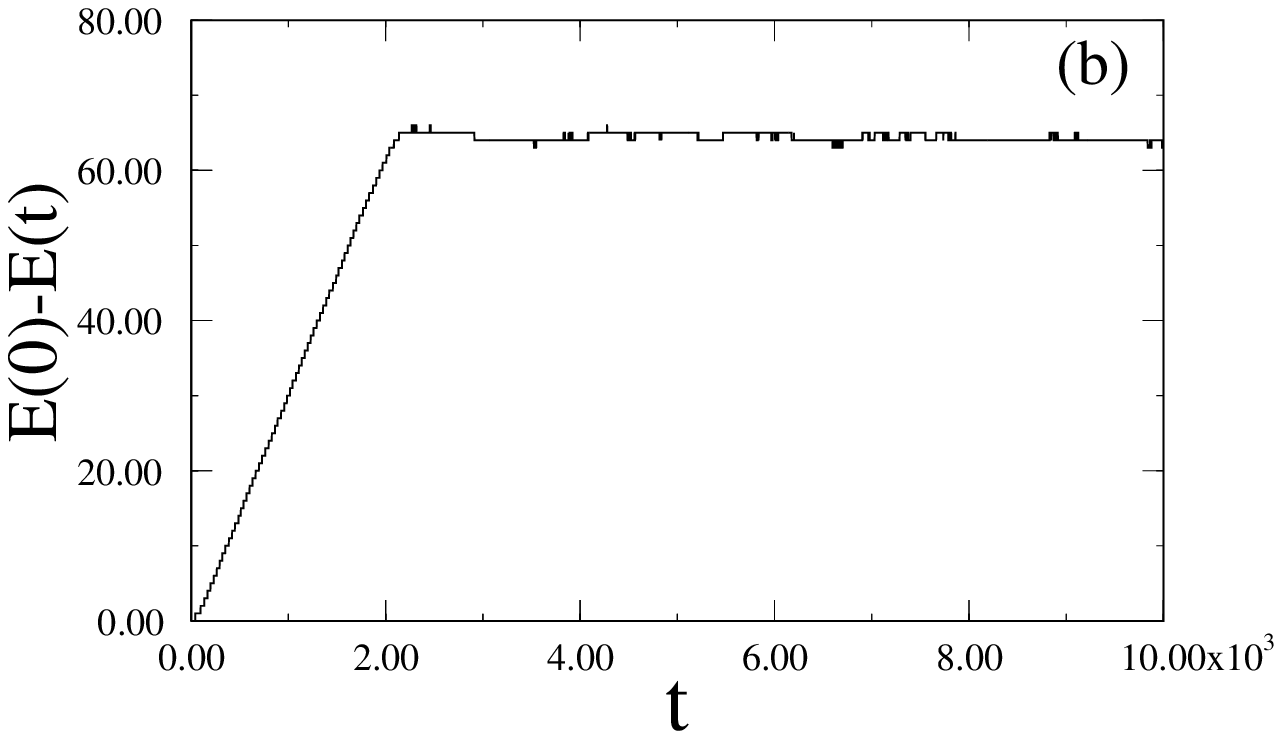}
\caption{
 (a) The density contrast  C(t) versus time step t. (b)
  Kinetic energy loss $E(0)-E(t)$ versus time step t. Here E(0) is the
  kinetic energy at t=0. The kinetic energy of a moving particle is
  chosen as  energy unit.
}
\end{figure}

\begin{figure}[e]
\vspace{5.9cm}
\includegraphics{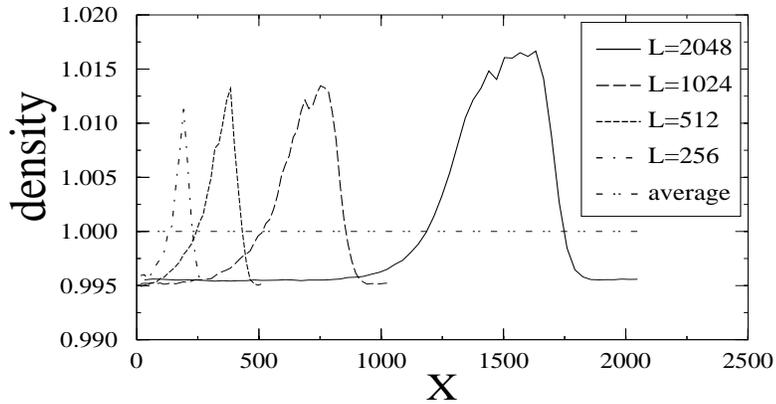}
\caption{
 Density as a function of position X along the pipe.
  The average has been made in the perpendicular direction. The model
  parameters are $\rho=1.0$, $p=0.1$, $g=0.5$, $b=0.5$. The width is
  fixed for various pipe lengths, $W=11$. }
\end{figure}

\begin{figure}[e]
\vspace{6.9cm}
\includegraphics{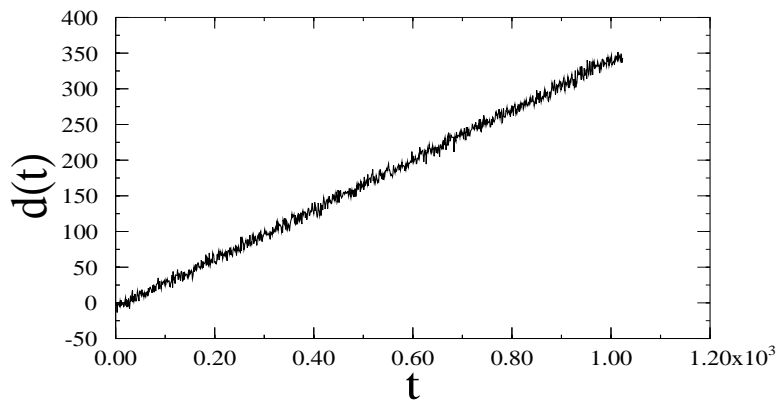}
\caption{ Real--space determination of the velocity of density
wave.  The horizontal
axis is the time interval while the vertical axis is the displacement
of the wave obtained by maximal overlap. The velocity is the slope of
the  line which is $0.36 \pm 0.05$ for a  system with $L=512$, 
$W=11$, $\rho=1.0$, $p=0.1$, $g=0.5$, $b=0.5$. }
\end{figure}

\begin{figure}[e]
\vspace{5.9cm}
\includegraphics{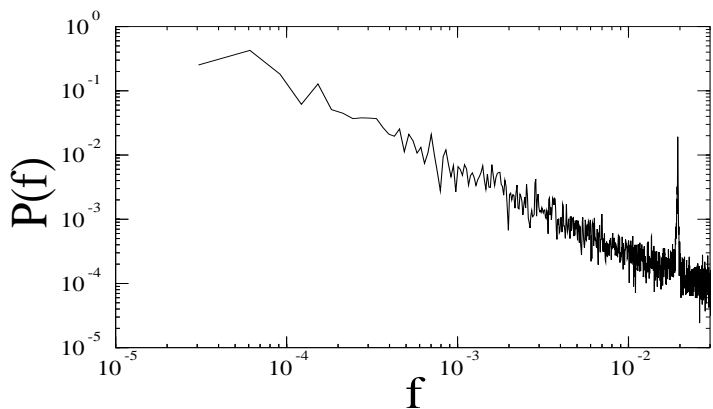}
\caption{ Power spectrum $P(f)$ of the time series of the density
fluctuation inside a region in a pipe of length L=220 and width W=11.
The model parameters are $p=0.8$, $b=0.5$, $g=0.5$, $\rho=1.0$. 
The time series of the density fluctuation were recorded every 10 time
steps 
and the time period correponding to a frequency $f$ is $10/f$.
 }
\end{figure}

\begin{figure}[e]
\vspace{6.9cm}
\includegraphics{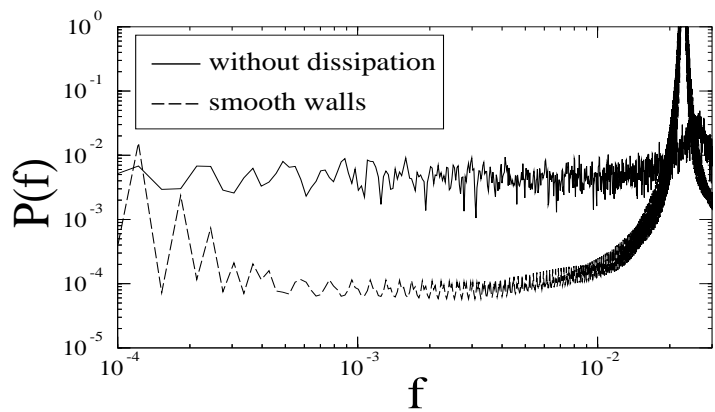}
\caption{ Power spectra $P(f)$ of the time series of the density
fluctuation inside a region in a pipe of length L=220 and width W=11.
 Either without dissipation or with smooth walls, white noise is
 observed. The model parameters for the system without dissipation are 
 $p=0$,  $b=0.5$, $g=0.5$, $\rho=1.0$ while $p=0.5$,  $b=0$, $g=0.5$,
 $\rho=1.0$ for the system with smooth walls.
 }
\end{figure}

\begin{figure}[e]
\vspace{5.9cm}
\includegraphics{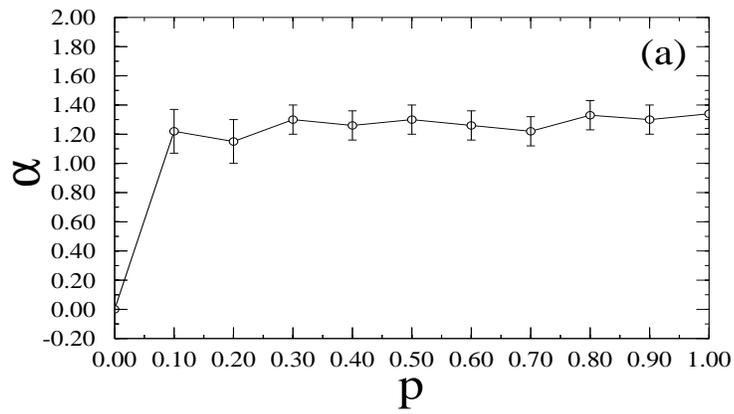}
\vspace{6.9cm}
\includegraphics{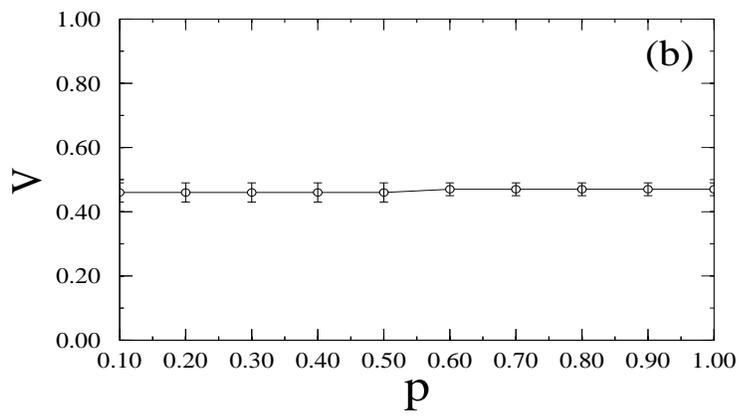}
\caption{ Dependence on the dissipation parameter $p$. The model
  parameters kept fixed are $b=0.5$, $g=0.5$ and $\rho=1.0$. (a) The
power--law decay exponent $\alpha$ of the power spectra. (b) The 
velocity of the density wave.
 }
\end{figure}

\begin{figure}[e]
\vspace{6.9cm}
\includegraphics{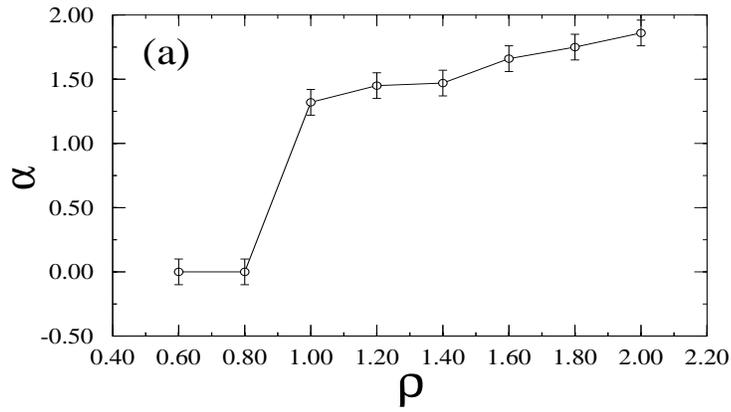}
\vspace{6.9cm}
\includegraphics{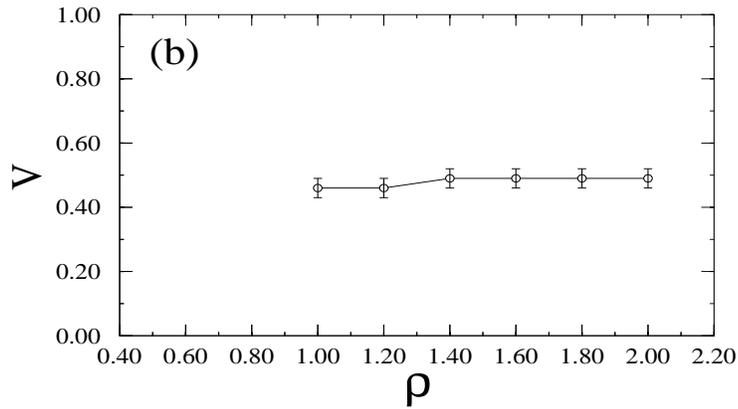}
\caption{ Dependence on the average density  $\rho$. The model
  parameters kept fixed are $p=0.5$, $b=0.5$ and $g=0.5$. (a) The
power--law decay exponent $\alpha$ of the  power spectra. (b) The 
velocity of the density wave.
 }
\end{figure}

\begin{figure}[e]
\vspace{5.9cm}
\includegraphics{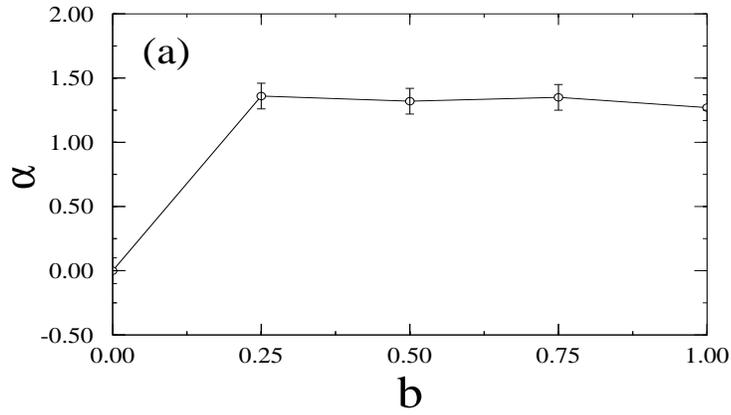}
\vspace{6.9cm}
\includegraphics{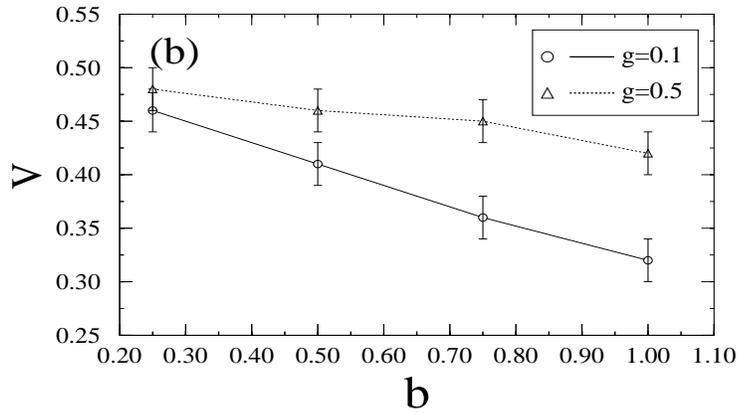}
\caption{ Dependence on the roughness  $b$ of the walls. (a) 
Power--law decay exponent $\alpha$ in the power spectra. The model
  parameters kept fixed are $p=0.5$, $g=0.5$ and $\rho=1.0$. (b)
  Velocity of the 
density wave for two different magnitudes of gravity. Here $p=0.5$ and
$\rho=1.0$. 
 }
\end{figure}

\begin{figure}[e]
\vspace{4.8cm}
\includegraphics{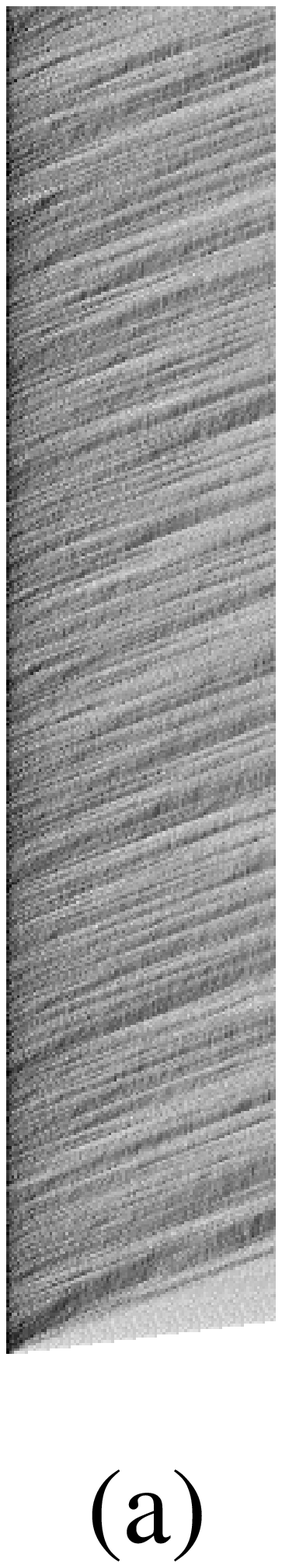}
\includegraphics{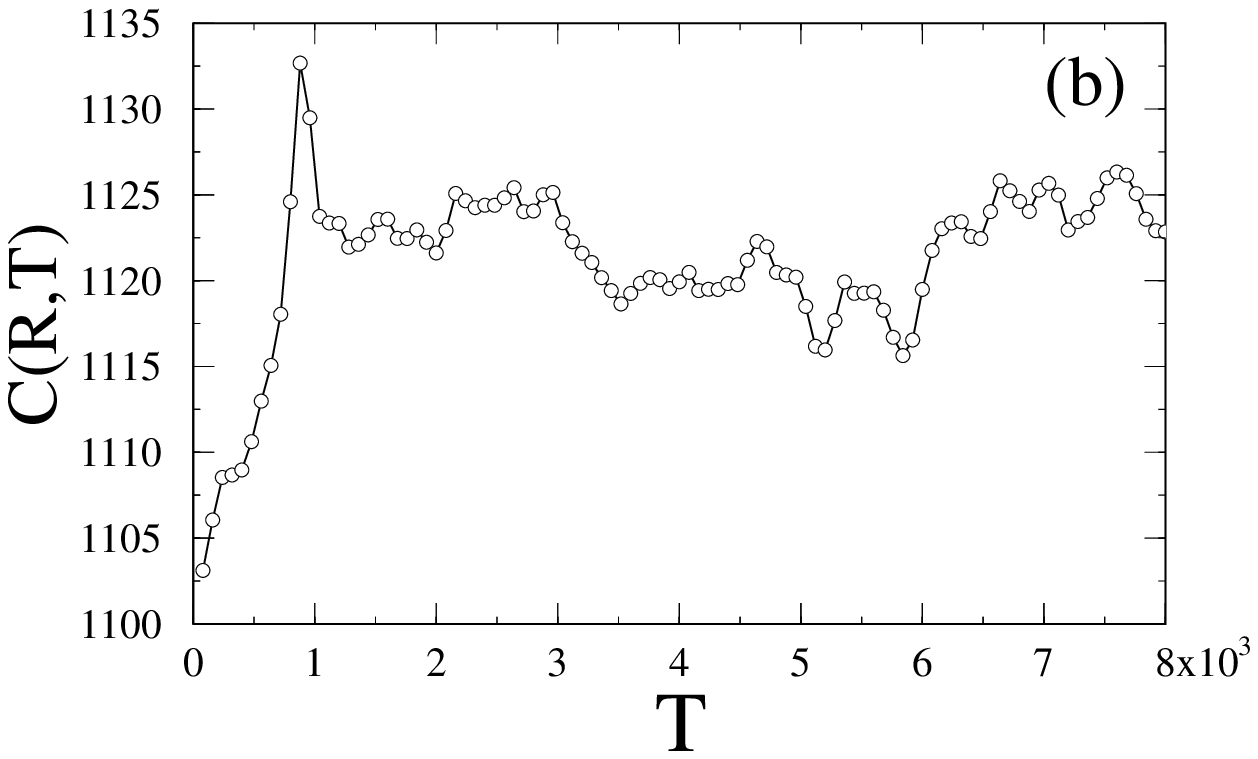}
\caption{ (a) Time evolution of the density $n_i \{i=1,2,\cdots 220\}$
  in the $100$ bins in the pipe of $L=1000$, $W=5$ and $I = 0.5$. Other
model parameters are $p=0.5$, $b=0.5$, $g=0.2$. 
 Densities at a given time are
plotted from left to right (direction of gravity) while
densities  at different time steps are plotted from bottom to top
 (direction of time increase).  Time goes from $0$ to $40,000$
 time steps. The  grayscale of each bin is a linear function of $n_i$.
Darker regions  correspond to higher densities. (b) Two--point
density--density correlation function $C(R,T)$ versus time difference
$T$ at a fixed spatial separation $R=30$ for the evolution shown in (a).
 }
\end{figure}

\begin{figure}[e]
\vspace{5.5cm}
\includegraphics{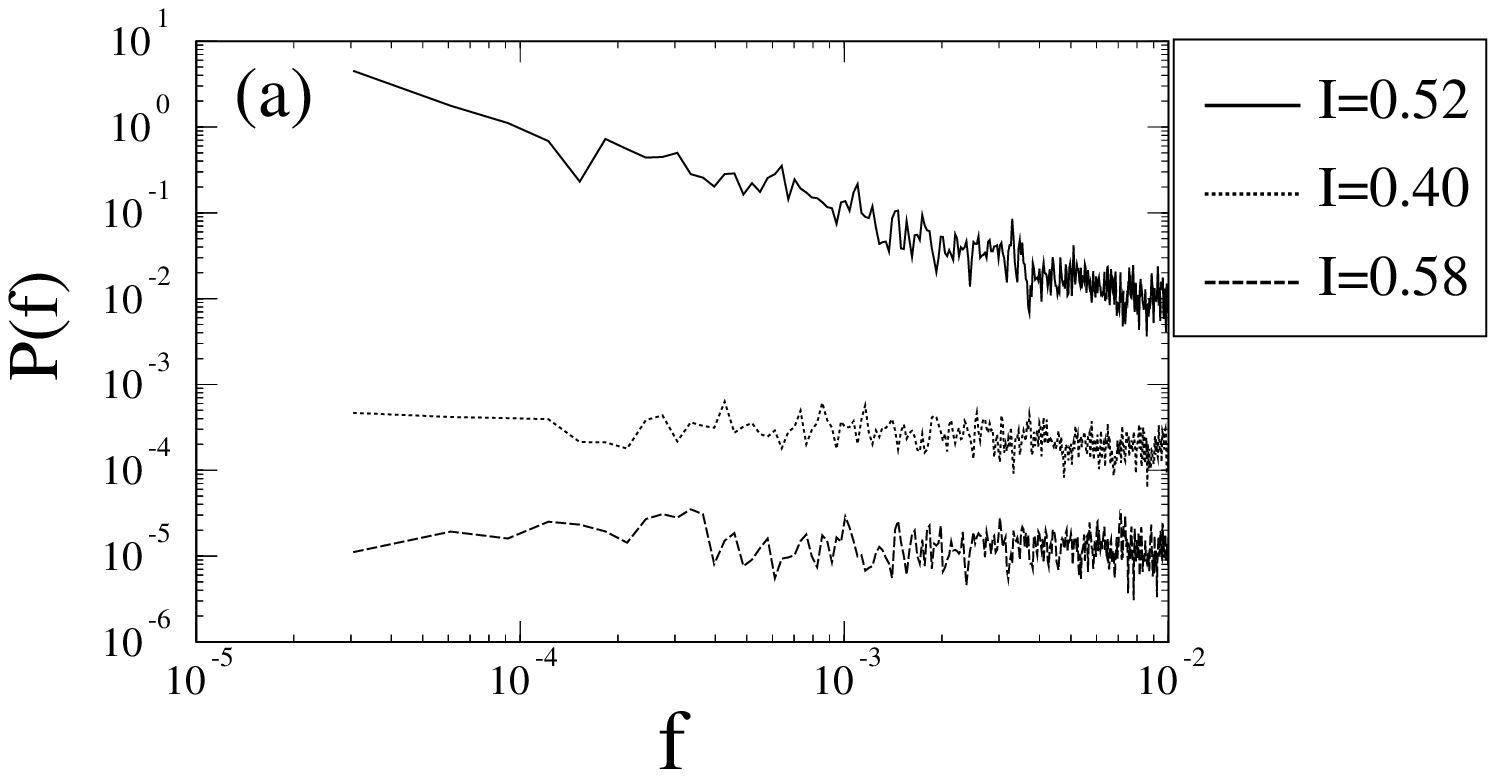}
\vspace{5.9cm}
\includegraphics{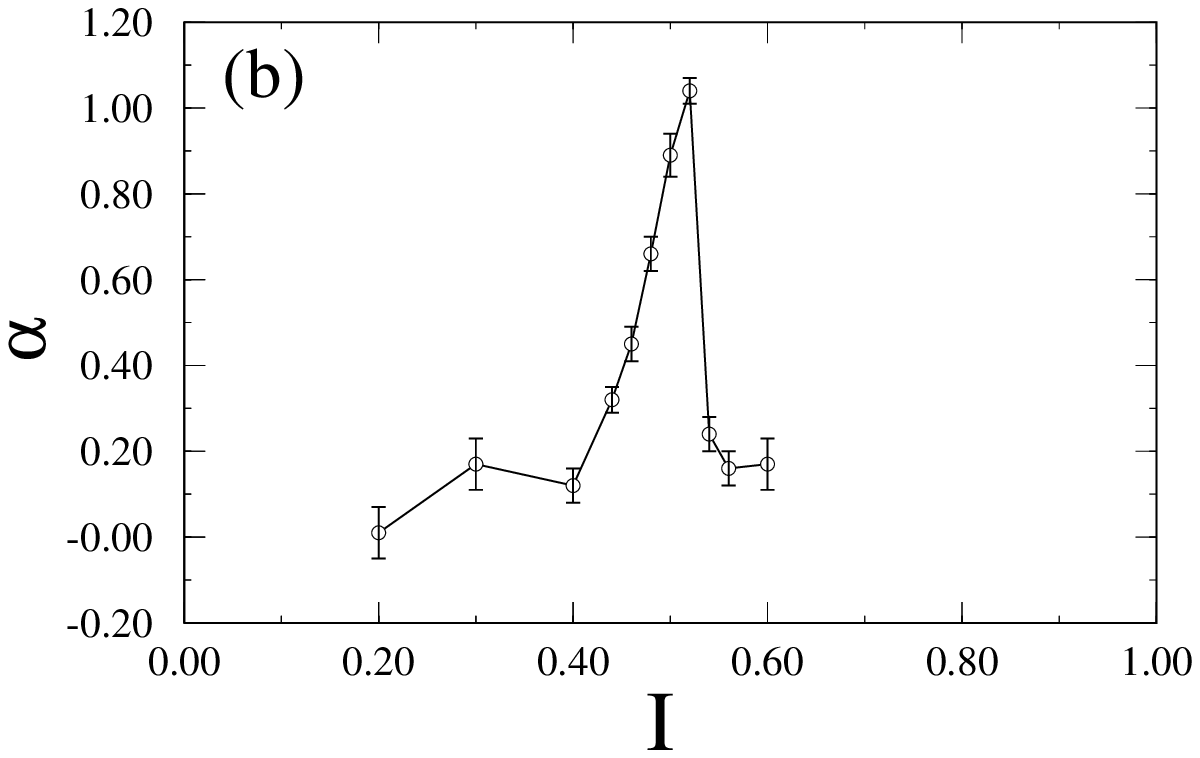}
\vspace{6.3cm}
\includegraphics{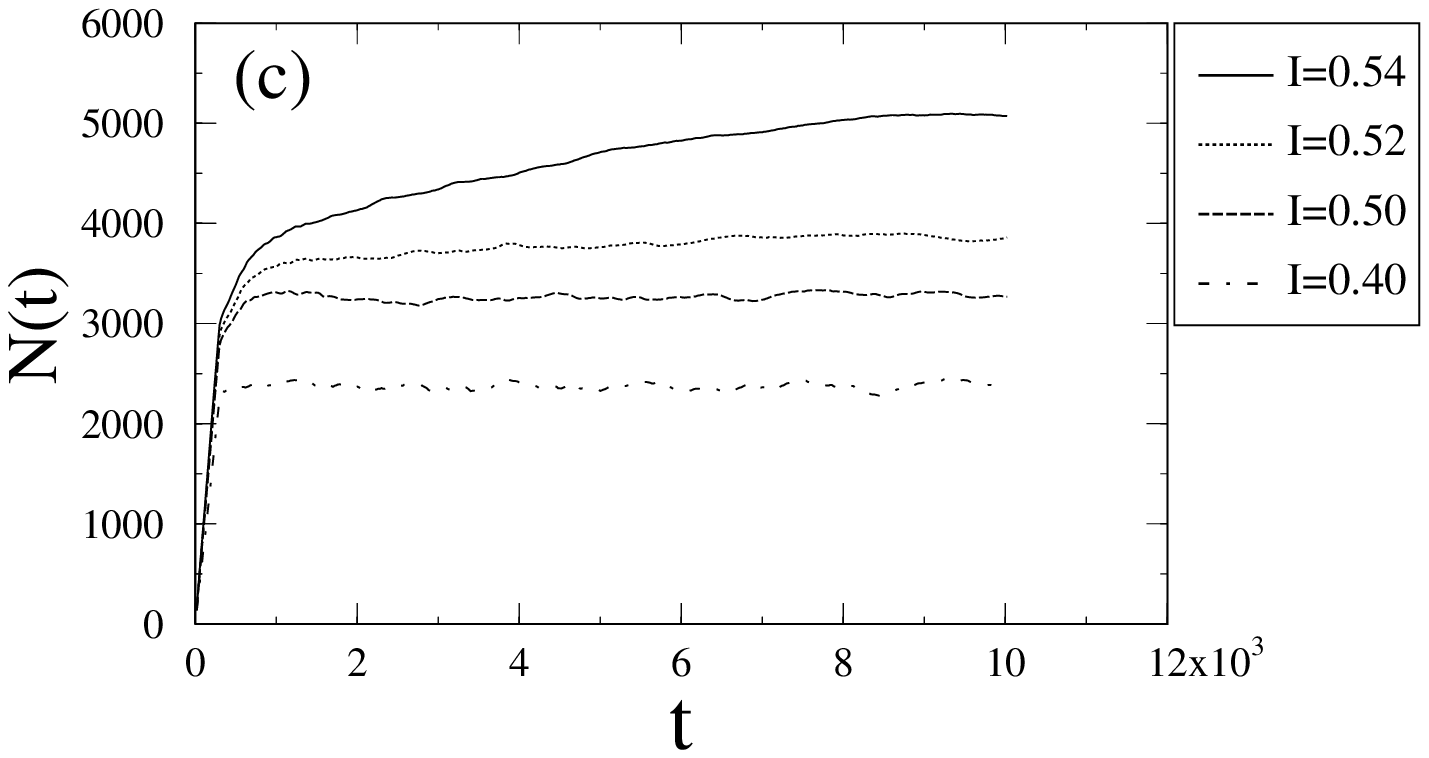}
\caption{ (a) Three typical power spectra for different injection
  rates $I$, $I=I_c=0.52, I < I_c, I > I_c$. The model parameters kept
  fixed are   $p=0.5$, $b=0.5$, $g=0.5$. The two lower curves have been
  shifted vertically for clarity.  (b) The exponent $\alpha$ in the
  power spectra $1/f^{\alpha}$ for different injection rates. (c)
  Total number of particles in the system $N(t)$ versus time step $t$
  for different injection rates.
  Each curve is an average over $32$ simulations. 
 }
\end{figure}

\newpage
\begin{thebibliography}{1}
\bibitem{Williams} J. C.~Williams, Powder Technol.{\bf 15}, 245
(1976).
\bibitem{Haff} P. K.~Haff and B. T.~Werner, Powder Technol.{\bf 48}, 239
(1986).
\bibitem{Rosato} A.~Rosato, K. J.~Strandburg, F.~Prinz, and R.
H.~Swendsen, Phys. Rev. Lett. {\bf 49}, 59
(1987).
\bibitem{Devillard} P.~Devillard, J. Phys. France {\bf 51}, 369
(1990).
\bibitem{Farady} M.~Faraday, Philos. Trans. R. Soc. London {\bf 52}, 299
(1831).
\bibitem{Evesque} P.~Evesque and J.~Rajchenbach, Phys. Rev. Lett. {\bf
62}, 44(1989).
\bibitem{Taguchi} Y. H.~Taguchi, Phys. Rev. Lett. {\bf
69}, 1367(1992).
\bibitem{Gallas} J. A. C.~Gallas, H. J.~Herrmann, and S.~Sokolowski,
Phys. Rev. Lett. {\bf 69}, 1371(1992).
\bibitem{Liu} C--h.~Liu and S. R.~Nagel,
Phys. Rev. Lett. {\bf 68}, 2301(1992).
\bibitem{Jaeger} H. M. Jaeger and S. R.~Nagel,
Science  {\bf 255}, 1523(1992).
\bibitem{nature} K.~L. Schick and A. A. Verveen, Nature {\bf 251},
599(1974).
\bibitem{Baxter} G. W.~Baxter, R. P.~Behringer, T.~Fagert, and G.
A.~Johnson, Phys. Rev. Lett.  {\bf 62}, 2825(1989).
\bibitem{Poschel} T.~P\"oschel, J. Phys. I France,  {\bf 4}, 499(1992).
\bibitem{Ristow} G.~Ristow and H. J.~Herrmann, Phys. Rev. E {\bf
50}, R5(1994).
\bibitem{Jysoo} J.~Lee, Phys. Rev. E {\bf 49}, 281(1994).
\bibitem{peter} P. Dimon, Private Communication.
\bibitem{shearflow} C.~S. Campbell and C.~E. Brennen, J. Fluid Mech.
  {\bf 151}, 167(1985); P.~A. Thompson and G.~S. Grest, Phys. Rev.
  Lett. {\bf 67}, 1751(1991); D.~M. Hanes and D.~L. Inman, J. Fluid Mech.
  {\bf 150}, 357(1985); O.~R. Walton and R.~L. Braun, J. Rheol. {\bf
    30}, 949(1986).
\bibitem{LeeHans} J.~Lee and H.~J. Herrmann, J. Phys. A {\bf 26}, 373(1993).
\bibitem{MC} A. Rosato, K.~J. Strandburg, F. Prinz and R.~H. Swendsen,
  Phys. Rev. Lett. {\bf 58}, 1038(1987); A.~D. Rosato, Y. Lan and
  D.~T. Wang, Powder Technol. {\bf 66}, 149(1991).
\bibitem{luding} S. Luding, E. Cl\'ement, A. Blumen, J. Rajchenbach
  and J. Duran, Phys. Rev. E {\bf 49}, 1634(1994).
\bibitem{CA} G.~W. Baxter and R.~P. Behringer, Phys. Rev. A {\bf 42},
  1017(1990); Physica D {\bf 51}, 465(1991).
\bibitem{MDbook} M.~P. Allen and D.~J. Tildesley, {\em Computer
    Simulations of Liquids}, Clarendon Press,
Oxford (1987).
\bibitem{FHP} U.~Frisch, B.~Hasslacher, and Y.~Pomeau, Phys. Rev. Lett. {\bf
56}, 1505(1986).
\bibitem{us} G. Peng and H. J. Herrmann, Phys. Rev. E {\bf 48}, R1796(1994).
\bibitem{Savage} S.~Savage, J. Fluid Mech. {\bf 241}, 109(1992).
\bibitem{Goldhirsch} I.~Goldhirsch and G.~Zanetti, Phys. Rev. Lett. {\bf
70}, 1619(1993).
\bibitem{flekkoy} E. Flekk\o y and H.~J. Herrmann, Physica A {\bf 199},
  1(1993).
\bibitem{book} W. H.~Press, B. P.~Flannery, S. A.~Teukolsky, and W.
T.~Vetterling, {\em Numerical Recipes in C}, Cambridge University Press,
Cambridge (1988).
\bibitem{Isarei} A. Bershadskii, preprint (1994).
\bibitem{wolfgang} W. Verm\"ohlen, G.~A. Kohring, S. Melin, H. Puhl
  and H.~J. Tillemans, HLRZ preprint,  {\bf 75/93},  (1993)
\bibitem{Jams} K.~Nagel and M.~Schreckenberg, J. Physique I {\bf 2},
2221(1992).
\bibitem{Kai} K.~Nagel, Int. J. Mod. Phys. C {\bf 5}, 567(1994). 
\bibitem{Janos} A. K\'arolyi and K. Kert\'esz, preprint, (1994).
\bibitem{Herrmann} H. J. Herrmann, in preparation. 

\end{thebibliography}
\end{document}